# Improving Short-Term Electricity Price Forecasting Using Day-Ahead LMP with ARIMA Models

Zhongyang Zhao, *Student Member, IEEE,* Caisheng Wang, *Senior Member, IEEE,* and Matthew Nokleby, *Member, IEEE*
Department of Electrical and Computer Engineering, Wayne State University, Detroit, U.S.
Carol J. Miller
Department of Civil and Environmental Engineering, Wayne State University, Detroit, U.S.

*Abstract*— Short–term electricity price forecasting has become important for demand side management and power generation scheduling. Especially as the electricity market becomes more competitive, a more accurate price prediction than the day-ahead locational marginal price (DALMP) published by the independent system operator (ISO) will benefit participants in the market by increasing profit or improving load demand scheduling. Hence, the main idea of this paper is to use autoregressive integrated moving average (ARIMA) models to obtain a better LMP prediction than the DALMP by utilizing the published DALMP, historical real-time LMP (RTLMP) and other useful information. First, a set of seasonal ARIMA (SARIMA) models utilizing the DALMP and historical RTLMP are developed and compared with autoregressive moving average (ARMA) models that use the differences between DALMP and RTLMP on their forecasting capability. A generalized autoregressive conditional heteroskedasticity (GARCH) model is implemented to further improve the forecasting by accounting for the price volatility. The models are trained and evaluated using real market data in the Midcontinent Independent System Operator (MISO) region. The evaluation results indicate that the ARMAX-GARCH model, where an exogenous time series indicates weekend days, improves the short-term electricity price prediction accuracy and outperforms the other proposed ARIMA models.

## I. INTRODUCTION

Load/demand management (also called demand response or demand side management) techniques have been important tools in improving the voltage profile, system efficiency and stability, and for matching the stochastic output power of renewable sources [1]. A majority of demand response programs are based on electricity price signals [2]. In addition to the programs following real-time price signals, there are also many demand response programs using price forecasts to do load scheduling. The publically available electricity price prediction is the day-ahead locational marginal price (DALMP) published by independent system operators (ISO) or Regional Transmission Organizations (RTOs). However, day-ahead and real-time are basically two different markets and there can be a large difference between the DALMP and real-time LMP. Hence, the motivation of this paper is to improve short-term forecasting on the order of a few hours, which is important for effective system operation and power demand management.

In time series analysis, ARIMA models are a great tool for forecasting a stationary time series and a non-stationary time series that can be made to be "stationary" by differencing and other possible transformation techniques. The Integrated (I) part in an ARIMA model is used to account for non-stationary elements in a time series. In this paper, a set of ARIMA models are developed for short-term LMP prediction. ARIMA models have been analyzed and evaluated for forecasting electricity price [3]-[6]. An ARIMA model is used to analyze the time series with Box and Jenkins method [11] and the next-day market clearing price (MCP) was predicted using the ARIMA model while considering explanatory variables, such as demand [3]. Furthermore, a wavelet transform was employed to decompose the ill-behaved historical price time series to a better-behaved constitutive series set [4]. After different ARIMA models were used for the better-behaved set, the inverse wavelet transform was then used to forecast the price [4]. A hybrid model using not only ARIMA but also artificial neural network (ANN) was used for electricity spot price forecast [5], [6]. A seasonal ARIMA (SARIMA) model was proposed to predict the time series for handling the weekly and daily periodical fluctuations [7], [8]. The SARIMA with exogenous data, such as power production [7] and temperature [8], was developed to forecast the day-ahead spot electricity price in Sweden by considering weekly seasonal effects [7] and to produce short-term PV generation forecasting [8].

Extensive work related to the electricity price forecast using ARIMA models has been completed [9], [10]. Nevertheless, the DALMP data has not been utilized to improve short-term real-time LMP prediction yet. In this paper, the DALMP is used instead of other market information because the DALMP published by ISOs is the best available information to represent the day-ahead market condition. As previously discussed, most of the system operation and demand side management are based on the DALMP information. However, the real-time nodal prices can fluctuate quite significantly due to unpredictable events such as transmission line and generator outages as well as the stochastic behaviors of customer load demands and renewable generation outputs. Hence, the DALMP is incorporated into ARIMA models to improve short-term prediction.

This paper first investigates the features of the DALMP and RTLMP data of MISO between January 2015 and December 2015 and forms a differential series between DALMP and RTLMP for increasing the forecasting accuracy. Four ARIMA models are then proposed and compared for electricity price forecasting in Section III. Meanwhile, a generalized autoregressive conditional heteroskedastic (GARCH) model is proposed to implement on the ARIMA models for the volatility present in the LMP series. In Section IV, the MISO LMPs between 1/4/2015 and 1/31/2016 are used to validate the proposed models. The results show that the ARMAX-GARCH model produces the best result for short-term electricity price prediction in the real-time electricity market. It can improve the forecasting accuracy by more than 27% for the one-hour-ahead predictions.

## II. DATA FEATURES

In this paper, two types of LMP data, namely the DALMP and RTLMP, are analyzed and utilized for improving the accuracy of short-term electricity price forecasting.

### A. Raw Data Ananlysis

The 24-h RTLMP between 1/5/2015 and 12/27/2015 in MISO is analyzed by observing the sample autocorrelation function (ACF) and partial autocorrelation function (PACF), shown in Fig. 1. The ACF and PACF plots are commonly used for ARIMA model selection using Box-Jenkins method [11]. The PACF plot is especially useful in identifying the order of an autoregressive model. In Fig. 1, a particular seasonal pattern is evident by the spikes appearing at approximately a 24-hour cycle. Hence, a SARIMA model should be investigated and tested to see whether a seasonal model might perform better. However, as discussed later in Section IV, a seasonal model does not necessarily guarantee a better performance if the seasonal pattern is weak.

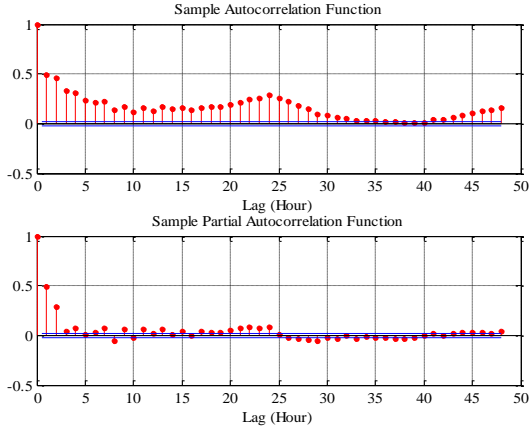

Fig. 1. ACF and PACF plots of the MISO RTLMP between 1/5/2015 and 12/27/2015.

### B. Differential Series between the DALMP and RTLMP

According to the Box-Jenkins method, the time series is assumed to be stationary in using ARMA model [11]. Hence, forming a stationary series is an important and necessary step before using an ARMA model.

The main idea of this paper is to utilize the DALMP data published in the day-ahead market by ISOs to improve the electricity price forecasting performance in short term. Both the historical RTLMP and DALMP series are used for improving the real-time LMP prediction. The series representing the differential between the past DALMP and RTLMP is extracted by (1):

$$\Delta LMP_t = DALMP_t - RTLMP_t, \quad t = 1, 2, \dots \quad (1)$$

where $\Delta LMP_t$ is the differential value between the DALMP and RTLMP at time $t$.

When $\Delta LMP_{t+i}$ is predicted by implementing the ARMA model, the forecasting $RTLMP_{t+i}$ is obtained by (2).

$$RTLMP'_{t+i} = DALMP_{t+i} - \Delta LMP_{t+i}, \quad i = 1, 2, \dots \quad (2)$$

where $DALMP_{t+i}$ is the DALMP published by the ISO for time $t+i$ and the $\Delta LMP_{t+i}$ is the forecasting differential LMP series at time $t+i$.

The stationarity of $\Delta LMP$ is revealed by the sample ACF and PACF plots in Fig. 2, because ACF decays quickly. Hence, the data of $\Delta LMP$ is appropriate to fit in the ARMA model. The three different LMP series, namely DALMP, RTLMP and $\Delta LMP$, introduced in this section, are to be used in the ARIMA models developed to forecast the electricity price in Section III.

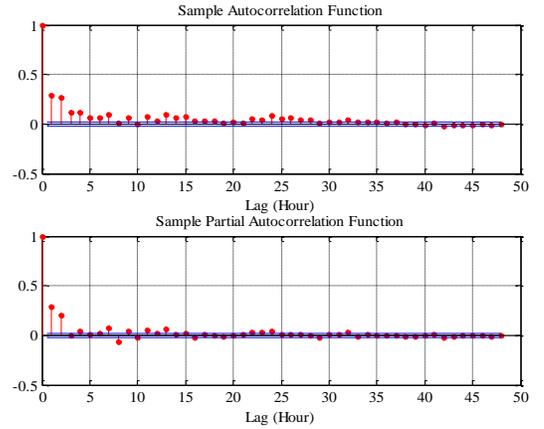

Fig. 2. ACF and PACF plots of $\Delta LMP$ of the MISO RTLMP between 1/5/2015 and 12/27/2015.

## III. MODEL DEVELOPMENT

Given the RTLMP with a seasonal pattern, a set of seasonal ARIMA models are proposed in this paper for short term forecasting, while an ARMA model is developed for $\Delta LMP$. Furthermore, exogenous data such as DALMP and weekday/weekend indicators are also incorporated into the ARIMA models to improve the forecasting accuracy.

### A. Seasonal ARIMA Model

The first model used in this paper is a seasonal ARIMA model. The seasonal ARIMA model for RTLMP is a standard $ARIMA(p,d,q) \times (P,D,Q)_s$ model described as follows [11], [12]:

$$\phi_p(B)\Phi_P(B^s)\nabla^d\nabla_s^D y_t = \mu + \theta_q(B)\Theta_Q(B^s)\varepsilon_t \quad (3)$$

where $B$ is the backward shift operator, i.e. $B^h y_t = y_{t-h}$; $p$ is the non-seasonal auto-regression (AR) order; $d$ is the non-seasonal differencing order; $q$ is the non-seasonal moving-average (MA) order; $P$ is the seasonal AR order; $D$ is the seasonal differencing; $Q$ is the seasonal MA order; and $S$ is the time span of repeating seasonal pattern. The error terms $\varepsilon_t$ are generally assumed to be the independent and identically distributed noise with zero mean and finite variance, i.e. Gaussian white noise; $\mu$ is a constant term.

$$\phi_p(B) = 1 - \phi_1(B) - \phi_2(B^2) - \cdots - \phi_p(B^p) \quad (4)$$

$$\Phi_P(B^s) = 1 - \Phi_1(B^s) - \Phi_2(B^{2s}) - \cdots - \phi_P(B^{Ps}) \quad (5)$$

$$\nabla^d = (1-B)^d \quad (6)$$

$$\nabla^D_s = (1-B^s)^D \quad (7)$$

$$\theta_q(B) = 1 - \theta_1(B) - \theta_2(B^2) - \cdots - \theta_q(B^q) \quad (8)$$

$$\Theta_Q(B^s) = 1 - \Theta_1(B^s) - \Theta_2(B^{2s}) - \cdots - \Theta_Q(B^{Qs}) \quad (9)$$

Furthermore, $\phi_1 \ldots \phi_p$, $\Phi_1 \ldots \Phi_P$, $\theta_1 \ldots \theta_q$ and $\Theta_1 \ldots \Theta_Q$ are the coefficients of autoregressive, seasonal autoregressive, moving average and seasonal moving average polynomials, respectively.

*B. ARMA Model*

Aiming at predicting the $\Delta LMP$ series, the ARMA model is used for estimating the electricity price in short-term. The $ARMA(p,q)$ model can be described as [11],[12]:

$$\phi_p(B) y_t = \mu + \theta_q(B) \varepsilon_t \quad (10)$$

where similar to the seasonal ARIMA model, $B$ is the backward shift operator; the $\phi_p(B)$, $\theta_q(B)$ and $\varepsilon_t$ have the same definitions as in the seasonal ARIMA model; $\mu$ is a constant term.

*C. Exogenous Variables*

Since the purpose of this paper is to improve the short term electricity price forecast compared to the DALMP which can be obtained from the ISO before the day, it is useful to accurately capture the relationship between DALMP and RTLMP. The regressor embedded into the SARIMA and ARMA models for handling the exogenous variables are given in the following SARIMAX model (11) and ARMAX model (12).

$$\phi_p(B)\Phi_P(B^s)\nabla^d\nabla^D_s y_t = \mu + \theta_q(B)\Theta_Q(B^s)\varepsilon_t + u'_t\gamma \quad (11)$$

$$\phi(B) y_t = \mu + \theta(B)\varepsilon_t + u'_t\gamma \quad (12)$$

with $\quad u'_t\gamma = \sum_{k=1}^{r} \gamma_k u_{tk} \quad (13)$

where $u'_t$ is the vector of exogenous variables $[u_{t1}\ u_{t2} \ldots u_{tr}]$ and $\gamma$ is the coefficients vector for exogenous variables, such as DALMP and weekday/weekend indices.

*D. Autoregressive GARCH Model*

Due to the high volatility and price spikes in the time series, models that allow for heteroskedastic data are needed. In the generalized autoregressive conditional heteroskedastic $GARCH(p,q)$ models [13]-[15], the conditional variance is dependent on the past values of the time series and a moving average of the past conditional variance, shown in (14) and (15):

$$\varepsilon_t = \sigma_t z_t \quad (14)$$

with $\quad \sigma_t^2 = \alpha_0 + \sum_{i=1}^{p}\alpha_i \varepsilon_{t-i}^2 + \sum_{j=1}^{q}\beta_j \sigma_{t-j}^2, \quad (15)$

where $\varepsilon_t$ is the error term or the residual return at time $t$; $z_t$ is basically a white noise process; $\alpha_0 > 0$, $\alpha_i > 0$ and $\beta_j > 0$.

Additionally, it is concluded in [15] that ARIMA-GARCH can outperform generic ARIMA when the time series is volatile. Hence, a GARCH model is also developed to deal with varying nodal electricity prices for a more accurate prediction.

*E. Evaluation Methods*

A percentage index $I_i$ is formulated in (16) for evaluating the performance of different prediction models. The percentage index $I_i$ displays how much the forecast accuracy improves when compared to the DALMP.

$$I_i = 1 - \frac{1}{n}\sum_{t=1}^{n}\left(\frac{|RTLMP_{t+i} - RTLMP'_{t+i}|}{|RTLMP_{t+i} - DALMP_{t+i}|}\right) \quad (16)$$

where $n$ is the sample number of the testing set; $i$ means the $i$-th prediction hour; $RTLMP_{t+i}$ is the real time LMP at time $t+i$; $RTLMP'_{t+i}$ is the $i$-th hour forecasting LMP at time $t+i$; $DALMP_{t+i}$ is the day-ahead LMP at time $t+i$;

The models developed and discussed in this section, namely SARIMA, ARMA, SARIMAX, ARMAX, ARMA-GARCH, and ARMAX-GARCH models will be evaluated and compared in the next section.

## IV. PREDICTION PERFORMANCE

The proposed forecasting models have been applied to predict the electricity prices in the region of MISO [16]. The training data set, from 1/5/2015 (Monday) to 12/27/2015 (Sunday), including 51 weeks hourly DALMPs and RTLMPs are used to obtain the parameters of the ARIMA models. The parameters estimation is based on the maximum likelihood estimation for the available training data. In this paper, the parameter estimation is accomplished by the estimate tool in MATLAB [17]. After the models are developed, the hourly RTLMP and DALMP data, from 1/4/2016 (Monday) to 1/31/2016 (Sunday), are employed for the model validation. The forecasting results show the developed models make more accurate short-term predictions than the DALMP provided by the ISO in the real-time market.

*A. Data Preprocessing*

A preprocessing scheme is proposed as (17) for eliminating large price spikes in the LMP series:

$$P_t = \begin{cases} UB\ \$/MWh & if\ RTLMP_t > UB\ \$/MWh \\ LMP_t & otherwise \\ LB\ \$/MWh & if\ RTLMP_t < LB\ \$/MWh \end{cases}, \quad (17)$$

where $UB$ and $LB$ are the upper bound and the lower bound and the values of $UB$ and $LB$ are defined according to the processed time series.

Furthermore, a natural logarithm transformation in [12], given in (18), is taken for suppressing larger fluctuations based on [12], because abnormal data points present in the observed time series could contribute to non-stationary and biased model fitting:

$$y_t = \log(P_t + c) \quad (18)$$

where $P_t$ is the LMP value and $c$ is a positive constant offset adding on $P_t$ to guarantee the logarithm transformation.

*B. SARIMA and SARIMAX Model Selection*

The minimum RTLMP value in the training set is $-28.7$\$/MWh, thus the offset $c$ in (18) is chosen to be 30 when RTLMP series from 1/5/2015 to 12/27/2015 is preprocessed by (18). The transformed series $y_{RT,t}$ can then be obtained. Based on the plots in Fig. 1, the seasonality in the SARIMA model is set to 24. As a result, the seasonal ARIMA model is established as $ARIMA(2,0,1) \times (1,1,1)_{24}$ in (19):

$$\phi_2(B)\Phi(B^{24})\nabla_{24}y_{RT,t} = \mu + \theta(B)\Theta(B^{24})\varepsilon_t. \quad (19)$$

In the SARIMAX model, the DALMP time series is used as the exogenous data for improving prediction performance.

Before incorporating the DALMP into the SARIMAX model, the DALMP series is preprocessed by (18) to obtain $y_{DA,t}$. Considering the minimum (negative) price in the DALMP time series and following the aforementioned RTLMP preprocessing, the offset $c$ for preprocessing the DALMP is also chosen to be 30 in (18). Hence, the SARIMAX model is determined as (20)

$$\phi_2(B)\Phi(B^{24})\nabla_{24}y_{RT,t} = \mu + \theta(B)\Theta(B^{24})\varepsilon_t + y'_{DA,t}\gamma \quad (20)$$

*C. ARMA and ARMAX Model Selection*

Based on the sample ACF and PACF plots of $\Delta LMP$ shown in Fig. 2, it is revealed that $\Delta LMP$ is stationary. Thus, the integrated part for reducing non-stationary term in the ARIMA model is not necessary for dealing with $\Delta LMP$. In other words, the initial differencing step is not required in the models for predicting the $\Delta LMP$.

Thus, ARMA and ARMAX models are developed to predict the $\Delta LMP$ series. Afterwards, the $RTLMP_{t+i}$ forecast can be obtained by (2).

Since the minimum value of $\Delta LMP$ series is -457.45 \$/MWh which is beyond a normal range, (17) is utilized to shape the series for a better model fitting. In (17), the $UB$ and $LB$ are set to 100 and -100, respectively. Then $P_{\Delta LMP,t}$ series is processed by (18) to become $y_{\Delta LMP,t}$ when the offset $c$ is set to 1000 for the transformation.

In addition, although the DALMP information has already been utilized in the prediction of $\Delta LMP$ when using the ARMA model, the additional weekday and weekend information is also valuable for consideration to further improve the prediction accuracy. In this paper, the weekday's hours are designed to be represented by "1" while weekend's hours are represented by "0". Therefore, a repeating 120 "1"s and 48 "0"s sequence (Note: 120 hours in the weekday and 48 hours in the weekend) is applied as the exogenous data to develop the ARMAX model.

Then Bayesian information criterion (BIC) [18] is utilized to handle the model order selection. As the BIC values of ARMA shown in Table I, the model order is selected to $ARMA(1,2)$ due to the lowest BIC value appearing at p=1 and q=2 in Table I. Similarly, the order of the ARMAX model is determined to be $ARMAX(1,1)$ according to the BIC values.

Table I
BIC VALUES OF THE ARMA MODEL

| p\q | 1 | 2 | 3 | 4 | 5 |
|---|---|---|---|---|---|
| 1 | -68875.1 | **-69085.5** | -69067.3 | -68905.7 | -69056.9 |
| 2 | -69073 | -69065.4 | -69082.9 | -68292.3 | -69008 |
| 3 | -69063.2 | -69050.2 | -68988.3 | -69069.4 | -69067.6 |
| 4 | -69084.7 | -69024.4 | -69042.1 | -69034.2 | -68958.5 |
| 5 | -69078.4 | -68944.7 | -69062.7 | -69031.2 | -69014 |

*D. Results Comparison*

When the $ARIMA(2,0,1) \times (1,1,1)_{24}$, $ARIMAX(2,0,1) \times (1,1,1)_{24}$, $ARMA(1,2)$ and $ARMAX(1,1)$ models are built, the GARCH model aiming at handling price volatility is applied. The $GARCH(1,1)$ model minimizes the BIC value, so it is implemented for implemented for every proposed model to account for volatility.

Table II
COMPARISON RESULTS OF ALL ARIMA MODELS

| Model | $I_1$ (%) | $I_2$ (%) | $I_3$ (%) |
|---|---|---|---|
| SARIMA | -7.09 | -23.16 | -29.59 |
| SARIMAX | -13.65 | -22.88 | -31.04 |
| ARMA | 26.64 | 13.03 | 6.79 |
| ARMAX | 26.50 | 12.75 | 6.53 |

Table III
COMPARISON RESULTS OF ALL ARIMA MODELS WITH $GARCH(1,1)$

| Model | $I_1$ (%) | $I_2$ (%) | $I_3$ (%) |
|---|---|---|---|
| SARIMA − GARCH | -6.10 | -20.66 | -27.27 |
| SARIMAX − GARCH | -29.53 | -57.55 | -66.52 |
| ARMA − GARCH | 27.14 | 16.85 | 10.70 |
| ARMAX − GARCH | 27.21 | 17.10 | 11.35 |

For validating and comparing the proposed models, the hourly LMP data set from 1/4/2016 (Monday) to 1/31/2016 (Sunday) is implemented. The short-term forecast results of every model are listed in Table II and Table III. In Table II, the improvement indices of the SARIMA and SARIMAX are negative and getting worse when the predicting hour increases, which means the SARIMA and SARIMAX models perform poorly compared to the DALMP series. The SARIMA and SARIMAX add two extra parameters to the model; the additional model complexity increases the risk of overfitting, but it does not help substantially in predicting RTLMP, which is only weakly seasonal.

In contrast, the one-hour-ahead prediction (i.e. $I_1$) percentage indices of the ARMA and ARMAX models are 26.6% better than the DALMP and while it is 13% better for the two-hour-ahead prediction indices. In other words, the ARMA models can provide a better prediction than the DALMP in short-term. The results of the different models with $GARCH(1,1)$ are listed in Table III, in which we can find the GARCH model is able to increase the prediction accuracy of the ARMA and ARMAX models by about 0.7% in the first hour and 3.5% in the second hour. Actually, the DALMP has predicted very well in January 2016 with mean absolute error (MAE) [19] of 2.06 \$/MWh while the MAE of the SARIMA, SARIMAX, ARMA and ARMAX with the GARCH model at the one-hour ahead prediction are 2.19 $/MWh, 2.67 $/MWh, 1.50 $/MWh and 1.50 $/MWh, respectively.

ARMA-GARCH and ARMAX-GARCH models have outperformed the other models on the next-hour and the next-couple-hour predictions. Both of them have done well and performed close to each other. For further comparing and analyzing these two models, the improvement indices $I$ of the next 12-hour prediction are plotted in Fig. 3.

According to Fig. 3, the ARMAX-GARCH beats ARMA-GARCH by about 0.9% after the 4th step prediction and it shows the model with exogenous variables could predict better than the one without the weekday/weekend index while both of the models have enough capability to outperform DALMP in the next 12-hour prediction. That means the inclusion of the weekday/weekend information as the exogenous variables is useful in improving the prediction accuracy in long term. At the same time, even though the forecasting capabilities of the models in Fig. 3 have little differences in first couple of hours, the models with $GARCH(1,1)$ demonstrate the advantages over a longer-term prediction.

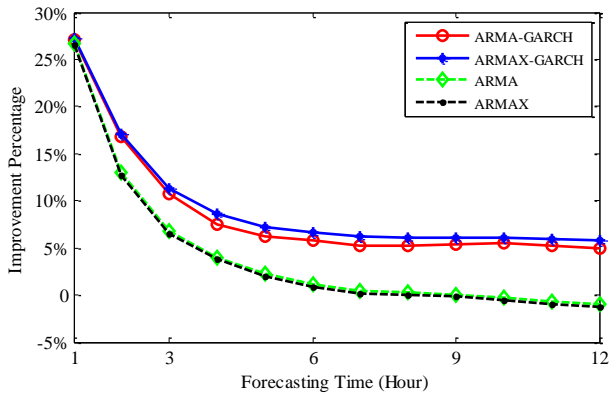

Fig. 3. Improvement indices of ARMA and ARMAX models in 12-hour prediction

## V. CONCLUSION

Various ARMA and ARIMA models (seasonal, non-seasonal, and with or without exogenous data included) were developed and compared in this paper for short-term electricity price prediction using the published DALMP and historical RTLMP data in an electricity market. A GARCH model was developed to handle price volatility for all the developed ARMA and ARIMA models. All the SARMA, SARMAX, ARMA and ARMAX models and the models with the GARCH model were developed and tested using the actual day-ahead and historical real-time LMPs data in MISO. The results show that adding seasonal data does not help improve the prediction accuracy. The results also show that the ARMAX-GARCH model with the weekday/weekend information as the exogenous data, has the best performance compared to all models. It achieves 27.21% improvement for the one-hour-ahead predictions, 17.10% improvement for the two-hour-ahead predictions and 11.35% improvement for the three-hour-ahead predictions. The ARMAX-GARCH model also shows over 5% improvement for predictions 12 hours ahead.

REFERENCES

[1] M. D. Ilic, L. Xie and J. Joo, "Efficient coordination of wind power and price-responsive demand - Part I: theoretical foundations," *IEEE Transactions on Power Systems*, vol. 26, no. 4, pp. 1875-1884, November 2011.

[2] G. Strbac, "Demand side management: Benefits and challenges," *Energy Policy*, vol. 36, pp. 4419–4426, December 2008.

[3] J. Contreras, R. Espínola, F. J. Nogales and A. J. Conejo, "ARIMA models to predict next-day electricity prices" *IEEE Transactions on Power Systems,* vol. 18, no. 3, pp. 1014-1020, August 2003.

[4] A. J. Conejo, M.A. Plazas, R. Espinola and A. B. Molina, "Day-ahead electricity price forecasting using the wavelet transform and ARIMA models," *IEEE Transaction on Power System*, vol. 20, no.2, pp. 1035-1042, May 2005.

[5] G. P. Zhang, "Time series forecasting using a hybrid ARIMA and neural network model," *Neurocomputing*, vol. 50, pp. 159-175, January 2003.

[6] R. Spokal, "Short-term hourly price forward curve prediction using neural network and hybrid ARIMA-NN model," *Information and Digital Technologies (IDT), 2015 International Conference*, Zilina, Slovakia, July 2015.

[7] M. Xie, C. Sandels, K. Zhu, L. Nordström, "A Seasonal ARIMA model with exogenous variables for elspot electricity prices in Sweden," *European Energy Market (EEM), 2013 10th International Conference*, Stockholm, Sweden, May 2013.

[8] S. I. Vagropoulos, G. I. Chouliaras, E. G. Kardakos, C. K. Simoglou and A. G. Bakirtzis, "Comparison of SARIMAX, SARIMA, modified SARIMA and ANN-based models for short-term PV generation forecasting," *Energy Conference (ENERGYCON), 2016 IEEE International*, Leuven, Belgium, April 2016.

[9] S. K. Aggarwal, L. M. Saini and A. Kumar, "Electricity price forecasting in deregulated markets: A review and evaluation," *International Journal of Electrical Power & Energy Systems*, vol. 31, Issue 1, pp. 13-22, January 2009.

[10] R. Weron, "Electricity price forecasting: A review of the state-of-the-art with a look into the future," *International Journal of Forecasting*, vol. 30, Issue 4, pp. 1030-1081, October-December 2014.

[11] G. E. P. Box, G. M. Jenkins and G. C. Reinsel, *Time Series Analysis: Forecasting and Control*, 4th Edition, Published by John Wiley & Sons, Inc., Hoboken, New Jersey, 2008.

[12] R. H. Shumway and D. S. Stoffer, *Time Series Analysis and Its Applications: with R Examples,* Third Edition, Springer New York Dordrecht Heidelberg London, 2011.

[13] R. Engle, "An Introduction to the Use of ARCH/GARCH models in Applied Econometrics," Available online: http://www.stern.nyu.edu/rengle/GARCH101.PDF, Access Date: November 2016.

[14] G. Ali, "EGARCH, GJR-GARCH, TGARCH, AVGARCH, NGARCH, IGARCH and APARCH Models for Pathogens at Marine Recreational Sites," *Journal of Statistical and Econometric Methods*, vol. 2, no. 3, pp. 57-73, 2013.

[15] R. C. Garcia, J. Contreras, M. van Akkeren and J. B. C. Garcia, "A GARCH forecasting model to predict day-ahead electricity prices," *IEEE Transaction on Power Systems*, vol. 20, Issue: 2, pp. 867-874, May 2005.

[16] Midcontinent Independent System Operator (MISO), Available Online: https://www.misoenergy.org/AboutUs/Pages/AboutUs.aspx, Access Date: November 2016.

[17] MATLAB: estimate, Available Online: https://www.mathworks.com/help/econ/arima.estimate.html, Access Date: November 2016.

[18] The Bayes Information Criterion (BIC), Available Online: http://www-math.mit.edu/~rmd/650/bic.pdf, Access Data: November 2016

[19] R. J. Hyndman and A. B. Koehler, "Another look at measures of forecast accuracy", *International Journal of Forecasting*, vol. 22, Issue 4, pp. 679-688, October-December 2006.